\documentclass[final,5p,times,preprint,twocolumn]{elsarticle}

\usepackage{amssymb}
\usepackage{lineno}
\usepackage{lipsum}
\usepackage{tabularx}
\usepackage{adjustbox}
\usepackage{arydshln}
\usepackage{amssymb, algorithm, algpseudocode}
\usepackage{graphicx}
\usepackage{pifont} 
\newcommand{\cmark}{\ding{51}}%
\newcommand{\xmark}{\ding{55}}%
\usepackage{adjustbox}
\usepackage{makecell}
\usepackage{colortbl}
\usepackage{array} 
\usepackage{bold-extra}  
\usepackage{xcolor}  
\usepackage{multirow}   
\usepackage{pagecolor}
\usepackage{amsmath}
\usepackage{hyperref}
\usepackage{float}
\usepackage{dsfont}
\usepackage{caption}
 \usepackage{booktabs}

\journal{Pattern Recognition Letters}

\begin{document}
% \linenumbers
\begin{frontmatter}

%% Title, authors and addresses

%% use the tnoteref command within \title for footnotes;
%% use the tnotetext command for theassociated footnote;
%% use the fnref command within \author or \address for footnotes;
%% use the fntext command for theassociated footnote;
%% use the corref command within \author for corresponding author footnotes;
%% use the cortext command for theassociated footnote;
%% use the ead command for the email address,
%% and the form \ead[url] for the home page:
%% \title{Title\tnoteref{label1}}
%% \tnotetext[label1]{}
%% \author{Name\corref{cor1}\fnref{label2}}
%% \ead{email address}
%% \ead[url]{home page}
%% \fntext[label2]{}
%% \cortext[cor1]{}
%% \affiliation{organization={},
%%             addressline={},
%%             city={},
%%             postcode={},
%%             state={},
%%             country={}}
%% \fntext[label3]{}

\title{Unpaired Translation of Chest X-ray Images for Lung Opacity Diagnosis via  Adaptive Activation Masks and Cross-Domain Alignment}
%% use optional labels to link authors explicitly to addresses:
%% \author[label1,label2]{}
%% \affiliation[label1]{organization={},
%%             addressline={},
%%             city={},
%%             postcode={},
%%             state={},
%%             country={}}
%%
%% \affiliation[label2]{organization={},
%%             addressline={},
%%             city={},
%%             postcode={},
%%             state={},
%%             country={}}
 
\author[label0]{Junzhi Ning \texorpdfstring{\corref{equal1}}}
\author[label3,label4]{Dominic Marshall \texorpdfstring{\corref{equal1}}}
\author[label0]{Yijian Gao}
\author[label2]{Xiaodan Xing}  
\author[label2]{Yang Nan}  
\author[label2]{Yingying Fang}  
\author[label2,label1]{Sheng Zhang}
\author[label4]{Matthieu Komorowski\texorpdfstring{\corref{equal2}}{}}
\author[label2,label1,label5,label6]{Guang Yang\texorpdfstring{\corref{equal2}}}
% \author[label2]{Matthieu Komorowski}
% \author[label3,label4,label5,label6]{Guang Yang}

% \cortext[equal1]{These authors contributed equally to this work.}
\cortext[equal1]{Co-first authors, contributing equally for this work.}
\cortext[equal2]{Co-last senior authors. Send correspondence to: Junzhi Ning (email: (j.ning23@imperial.ac.uk) and Guang Yang (email: g.yang@imperial.ac.uk)}

\affiliation[label0]{
    organization={Department of Computing, School of Engineering, Imperial College London}, % Department and Organization 
    city={London},
    postcode={SW7 2AZ},
    country={UK}
}

\affiliation[label3]{organization={Cleveland Clinic London},%Department and Organization
country={UK}}

\affiliation[label4]{organization={Department of Surgery and Cancer, Imperial College London},
            city={London}, 
%Department and Organization
            country={UK}}

\affiliation[label2]{organization={Bioengineering Department and Imperial-X, Imperial College London},%Department and Organization
            city={London},
            postcode={W12 7SL}, 
            country={UK}}

\affiliation[label1]{organization={National Heart and Lung Institute, Imperial College London},%Department and Organization
            city={London},
            postcode={SW7 2AZ},  
            country={UK}}

\affiliation[label5]{organization={Cardiovascular Research Centre, Royal Brompton Hospital},%Department and Organization
            city={London},
            postcode={SW3 6NP},  
            country={UK}}

\affiliation[label6]{organization={School of Biomedical Engineering \& Imaging Sciences, King's College},%Department and Organization
            city={London},
            postcode={WC2R 2LS},  
            country={UK}}
\begin{abstract}
%% Text of abstract
Chest X-ray radiographs (CXRs) play a pivotal role in diagnosing and monitoring cardiopulmonary diseases. However, lung opacities in CXRs frequently obscure anatomical structures, impeding clear identification of lung borders and complicating localization of pathology. This challenge significantly hampers segmentation accuracy and precise lesion identification, crucial for diagnosis. To tackle these issues, our study proposes an unpaired CXR translation framework that converts CXRs with lung opacities into counterparts without lung opacities while preserving semantic features. Central to our approach is the use of adaptive activation masks to selectively modify opacity regions in lung CXRs. Cross-domain alignment ensures translated CXRs without opacity issues align with feature maps and prediction labels from a pre-trained CXR lesion classifier, facilitating the interpretability of the translation process. We validate our method using RSNA, MIMIC-CXR-JPG and JSRT datasets, demonstrating superior translation quality through lower Fréchet Inception Distance (FID) and Kernel Inception Distance (KID) scores compared to existing methods (FID: 67.18 vs. 210.4, KID: 0.01604 vs. 0.225). Evaluation on RSNA opacity, MIMIC acute respiratory distress syndrome (ARDS) patient CXRs and JSRT CXRs shows our method enhances segmentation accuracy of lung borders and improves lesion classification, further underscoring its potential in clinical settings (RSNA: mIoU: 76.58\% vs. 62.58\%, Sensitivity: 85.58\% vs. 77.03\%; MIMIC ARDS: mIoU: 86.20\% vs. 72.07\%, Sensitivity: 92.68\% vs. 86.85\%; JSRT: mIoU: 91.08\% vs. 85.6\%, Sensitivity: 97.62\% vs. 95.04\%). Our approach advances CXR imaging analysis, especially in investigating segmentation impacts through  image translation techniques. 

\end{abstract}

% \end{highlights}
\begin{keyword}
%% keywords here, in the form: keyword \sep keyword

%% PACS codes here, in the form: \PACS code \sep code

%% MSC codes here, in the form: \MSC code \sep code
%% or \MSC[2008] code \sep code (2000 is the default)

% Medical Image Synthesis \newline 
Unpaired CXRs Translation \sep Lung Disease Decomposition \sep Lung Segmentation

\end{keyword}

\end{frontmatter}

%% \linenumbers

%% main text 
\section{Introduction}
The chest radiograph (CXR) is the most common radiological investigation and is widely performed across a range of medical settings for diagnosing and monitoring a variety of cardiopulmonary diseases. Compared to Computed Tomography (CT) scanning, CXRs are readily accessible, inexpensive, and result in minimal radiation exposure. Along with the digitisation of medical image analysis, analysing CXRs with algorithm-aided approaches has become an increasingly indispensable tool in the modern diagnostic toolkit \cite{adelaja2023operating}.  Despite the demonstrated efficacy and promising clinical applications, most computer vision approaches fail to accurately segment and detect the lung regions, generating a mask that captures the areas of the CXR that are relevant for the specific downstream clinical diagnostic task, particularly when faced with CXRs of varying technical quality or pathological cases, such as rotation, positioning, penetration and the presence of pulmonary diseases. 

\begin{figure}[!htb]
    \centering
    \includegraphics[width=0.9\linewidth]{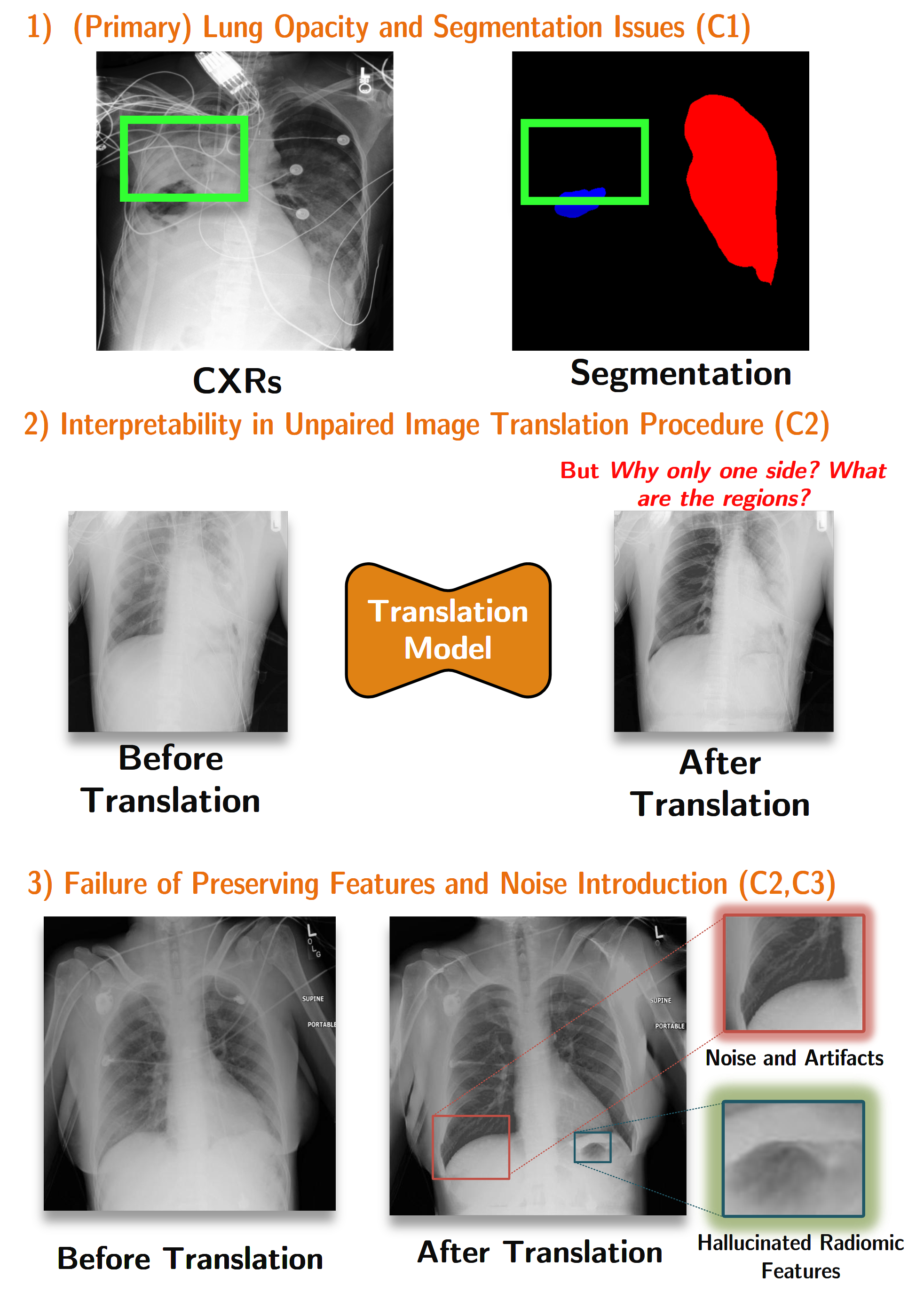}
    \caption{\textbf{Challenges identified in both lung segmentation with lung opacities and the unpaired image translation approach.} 1) Difficulty in correctly segmenting lung regions given CXRs with lung opacities. 2) Interpretability of the translation process when unexpected translation results occur. 3) Introducing unnecessary noise and artifacts. The C1, C2, C3 indicate the numbering from the contributions.}
    \label{fig:motivation} 
\end{figure}

To alleviate these issues, an unpaired CXR translation model can be utilized to efficiently transform CXRs with heavy lung opacities into ones without, providing clinicians with clearer visualization of lung regions. After translation, the enhanced visualization aids in better disease diagnosis and monitoring, particularly when pathological changes obscure normal anatomical landmarks or when visual information about anatomical structures is difficult to identify.  While unpaired image translation has been explored in the medical domain, current research on unpaired CXR translation still faces challenges, which are illustrated in \autoref{fig:motivation}: One challenge is the loss of semantic information in the translation impeding correct lung segmentation before and after translation for CXR analysis tasks. Additionally, the interpretability remains unclear regarding why certain regions are selectively modified while others are not, and which anatomical areas are specifically targeted. The third issue involves preserving the original anatomical features from the initial domain during translation, which is critical for maintaining clinical relevance in diagnosis and treatment planning.

Motivated by these challenges, our study primarily focuses on enhancing interpretability and preserving critical anatomical features through deep-learning models designed for lung opacity removal via unpaired CXR translation. We employ an RSNA dataset of chest X-rays with pulmonary opacities, to train the unpaired translation model to remove the lung opacities and assess its performance across CXR datasets including cases of acute respiratory distress syndrome (ARDS). ARDS is a lung pathology frequently encountered in intensive care units where patients develop acute respiratory distress and low blood oxygen levels due to extensive pulmonary involvement. Additionally, to improve interpretability, we propose directly modelling the activation masks that supervise the translation process and determine which regions need to be preserved or changed. By incorporating refined penalty terms during mask learning, our model generates better effective masks, ultimately enabling faithful visualisation of the CXR regions that are undergoing actual change during translation. Furthermore, in the absence of direct supervision in the unpaired CXR translation, we utilise priors from a pre-trained external classifier to provide alignment at feature and label levels. The alignments help correct the artifacts and preserve the original radiomic features in the chest radiographs. {\color{black} By addressing these challenges, the removal of lung opacities via CXR translation can enhance the delineation of lung boundaries, thereby improving the overall quality of CXRs for identifying pulmonary pathologies and supporting clinical decision-making for personalized treatment. Moreover, the effective analysis of lung opacities on individual CXRs can expand the diagnostic information derived from CXRs, which can contribute to the current trend of integrating multiple modalities of diagnostic data for clinical reference.}
 
{\color{black} In summary, this work makes the following contributions: 
(C1) We propose an unpaired image translation model that converts CXRs with lung opacities to normal CXRs, providing a healthy template to enhance and maintain lung segmentation performance on both in-distribution and out-of-distribution datasets respectively. (C2) We incorporate an activation mask mechanism with enhanced penalty terms for selective feature transformation, improving model interpretability by highlighting specific regions of change during the translation process. (C3) We design a cross-domain alignment module to minimize visual artifacts and preserve real CXR characteristics, which is validated through extensive downstream CXR analysis tasks demonstrating both enhanced reliability and maintained diagnostic utility for clinical applications. }

\begin{figure*}[!htb]
    \centering
    \includegraphics[width=1\linewidth]{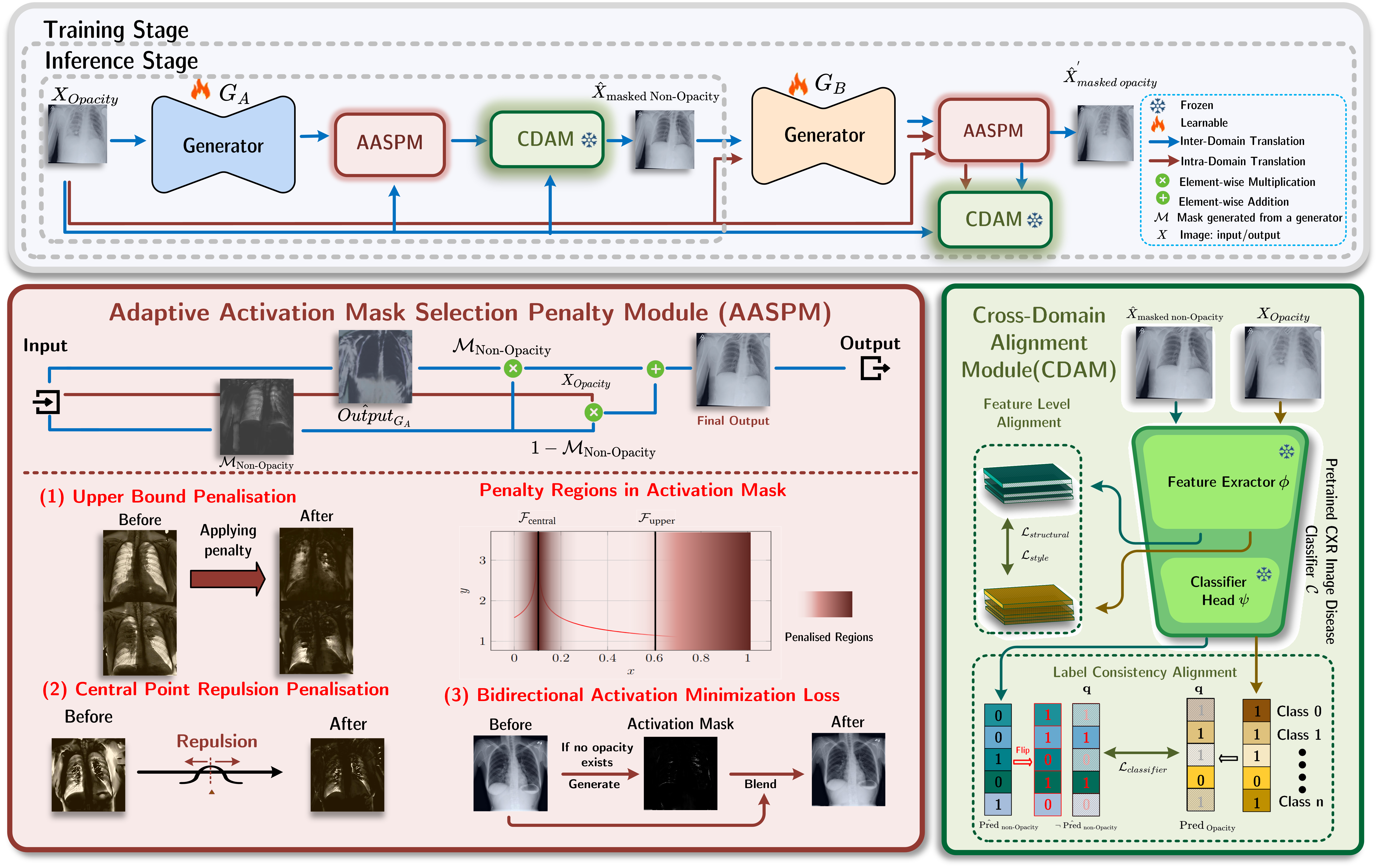}
    \caption{
    \textbf{Top: Model Overview of Proposed Method.}To eliminate the opacity parts in the CXRs, both CXRs with and without lung opacities were trained in a cycle using two domains and parameters of the generators that were learned for each direction. During the inference stage, only \textbf{Generator A} is used to generate the output of the CXRs. \textbf{Bottom Right:} \textbf{Illustration of Cross-Domain Alignment Module (CDAM).} Two types of alignment are present in this module: feature alignment and label consistency alignment. \textbf{Bottom Left:} \textbf{Graphical Illustration of AASP Module.} Two types of activation mask penalties and one minimization loss are used to control the generation of the activation masks. {\color{black} \textit{Intra-domain} translations refer to cases where the input CXR domain matches the generator's output domain (e.g., A~\(\rightarrow\)~A), whereas \textit{inter-domain} translations occur when the input CXR domain differs from the generator’s output domain (e.g., A~\(\rightarrow\)~B).}
    }
    \label{fig:Model_Overview}
\end{figure*}

\section{Related works} 
\subsection{Unpaired Image-to-Image Translation} 
While image translation in paired datasets has demonstrated its usefulness in various applications, many application domains have limited access to the paired datasets to facilitate the paired image translation models. Built upon Pix2pix \citep{isola2017image}, CycleGAN\cite{zhu2017unpaired} establishes the idea of the loss function design in the absence of the paired datasets through a cycle consistency constraint. The consistency of the image before and after reconstruction provides supervised guidance and relieves the need for paired datasets on the target domain. Similar work incorporating this cycle consistency is introduced in \cite{yi2017dualgan, kim2017learning}. In \cite{liu2017unsupervised}, UNIT relies on the assumption of the shared latent space that is common between the images conveying similar content on top of the cycle consistency of the latent embeddings. In 2018, MUNIT \cite{huang2018multimodal} assumes that the image space can be decomposed into embeddings for domain-invariant (content info) and domain-specific (style info) features without consistency loss. The Drit model from \cite{lee2018diverse}, furthermore, extends the idea of modelling the latent spaces specified for the content and style features in combination with the cross-cycle consistency constraints.  

\subsection{Medical Image Translation and Applications in CXRs} 
Image translation methods have also been applied in chest X-ray analysis tasks to assist or enhance the accuracy and performance of pathology detection and classification problems. \cite{tang2019abnormal} proposes a method of abnormal Chest X-ray detection through training one additional conditional generative model. The role of the conditional generative model is to reconstruct the normal Chest X-rays and learn a reduced latent space to reconstruct the given normal Chest X-rays. Only normal chest X-rays are supplied as the only source of training data. When presented with abnormal lung images during the testing phase, the reconstruction quality of the abnormal Chest X-rays is lower compared to the normal lung CXRs. The discrepancy in the reconstruction quality reveals the knowledge to identify the Chest X-rays as normal and abnormal. The method proposed in \cite{kim2023abnormality} designs a multi-stage deep learning model to achieve abnormality detection in CXRs. The main idea is to first develop a mechanism to find the matching pair of an abnormal lung image using a k-nearest neighbour method within a dataset to establish a paired dataset of abnormal-to-normal lung images. Then, the created paired dataset is used to train the paired image translation model. These results show that the residual map is computed based on the difference between the generated normal lung image and the abnormal lung image, improving the detection accuracy. \cite{tang2021disentangled} also exploits a similar idea of the residual maps through adversarial unpaired image translation to decompose the lesion regions via three branches of network modeling. The resulting residual maps are employed to improve the diagnostic performance of classification results compared to the abnormal lung X-rays.

\section{Method}
{\color{black}As illustrated in \autoref{fig:Model_Overview}, our framework consists of the following modules: two generators, one for each direction of the image translation task. After the generators' outputs, an Adaptive Activation Mask Selection Penalty Module is applied. The translated CXRs are then aligned with a pre-trained disease classifier at feature and prediction label levels within the target domain following translation. The gradient information will be backpropagated to update the parameters of the generators.}

{\color{black}
\subsection{Problem Notations}
In \autoref{fig:Model_Overview}, the generators $G_A$ and $G_B$ enable translation between CXRs with and without lung opacities. Inputs are $X_{\text{Opacity}} \in \mathcal{D}_{\text{Opacity}}$ and $X_{\text{Non-Opacity}} \in \mathcal{D}_{\text{Non-Opacity}}$, both with shape $C \times H \times W$. A hat ($~\hat{ }~$) denotes translated outputs, while a prime ($~^{\prime}~$) distinguishes different output types. Although translation is bidirectional in unpaired settings, we illustrate our approach using the translation from the CXR domain with lung opacities to the domain without lung opacities. The same notation applies to the reverse direction.
}

\subsection{Adaptive Activation Mask Selection Penalty Module (AASPM)}
    AASPM is designed to prevent the over-activation of the mask by the generators during the translation process. {\color{black}Unlike other CXR translation methods, where changes during the translation are revealed through direct computation of differences in model results or other visualisation techniques, we aim to leverage direct mask modelling selection mechanisms to generate model outputs dynamically. The activation mask, one of the outputs of the model's generator, serves a dual purpose: 1) it provides a visual overview of the areas undergoing transformation and 2) it elucidates the model's decision-making process in determining which regions of the CXRs require preservation of original features and which necessitate synthetic generation with better explainability.} Specifically, as illustrated in the lower left section of \autoref{fig:Model_Overview}, we focus on the translation process from the domain of lung opacities to the domain without lung opacities. The generator $G_A$ of the model produces two outputs, $\hat{Output}_{G_A} \in \mathbb{R}^{h \times w}$ and $\mathcal{M}_{\text{Non-Opacity}} \in \mathbb{R}^{h \times w}$: A pure model estimate of pixel values called $\hat{Output}_{G_A}$ is used to remove the opacity areas from the initial CXRs, and $\mathcal{M}_{\text{Non-Opacity}}$ is the activation mask that adaptively selects regions blended from $\hat{Output}_{G_A}$ and $X_{\text{Opacity}}$. This selection can be formulated as:
\begin{equation}
\resizebox{0.44\textwidth}{!}{$
    \hat{X}_{\text{masked
    Non-Opacity}} =  \hat{\text{Output}}_{G_A} \cdot \mathcal{M}_{\text{Non-Opacity}} + (1 - \mathcal{M}_{\text{Non-Opacity}}) \cdot X_{\text{Opacity}}, $}
\end{equation}
{\color{black} \noindent where the normalization of $\mathcal{M}_{\text{Non-Opacity}}$ is first performed using the min-max method, mapping values from [-1,1] to [0,1].}
\paragraph{\textbf{Upper Bound Penalisation}}
To preserve a level of coverage for the activation mask and to minimize the introduction of unnecessary artificial visual features from the model output, we constrained the activation mask by penalising the loss as follows: For each image $X_{\text{Opacity}} \in \mathcal{D}_{\text{Opacity}}$ in the dataset,

\begin{equation}
\mathcal{P}_{upper} = \operatorname{ReLU}\left(\sum_{m \in \mathcal{M}_{\text{Non-Opacity}}} m - \mathcal{F}_{upper}\right)^{2}.
\end{equation}

\noindent In our case, $m$ represent the activation value at each pixel from $\mathcal{M}_{\text{Non-Opacity}}$ and we choose the value of the $\mathcal{F}_{upper}$ to be 0.75. 

\paragraph{\textbf{Central Point Repulsion Penalisation}} The soft penalisation of the repulsion of the central point was introduced to encourage the activated map to not be located around the point $\mathcal{F}_{central}$ at an individual pixel level. The intuition was to encourage the model generator to produce the activation masks ranging away from the $\mathcal{F}_{central}$ so that the activation values were either significantly above or below a certain threshold, delivering the area of activation appropriately for opacity regions. To control the sensitivity of this loss function behaviour, we employed the hyperparameter of $\mathcal{F}_{height}$ to define the sensitivity of this repulsion.  Mathematically, this could be formulated as: for each image $X_{\text{Opacity}} \in \mathcal{D}_{\text{Opacity}}$ dataset,

\begin{equation}
    \mathcal{P}_{repul}  = \sum_{m \in \mathcal{M}_{\text {Non-Opacity }}}\left(\frac{1}{\left|m-\mathcal{F}_{\text {central }}\right|+\epsilon}\right)^{\mathcal{F}_{\text {height }}}
\end{equation}
where the value of $\epsilon$ was set to be 0.01 to avoid the zero-division error. For the training of the generators, the values of $\mathcal{F}_{central}$ and $\mathcal{F}_{height}$ are set to be 0.1 and 0.2, respectively.

\paragraph{\textbf{Bidirectional Activation Minimisation Loss (BAML)}}
{\color{black}
The translation generators in the framework can easily become overfitted to specific directions of translation, from one specific domain to another, and may produce over-activated activation masks regardless of the input domain. This tendency undermines the meaningful interpretation of the activation mask used to explain the lesion regions during translation. A bidirectional activation minimisation loss is leveraged for both $G_A$ and $G_B$ generators to help discriminate the difference between CXR domains. During training, the $\mathcal{L}_{1}$ loss minimises the pixel values of the activation masks to ensure that the generator learns to recognise the CXRs that do not require translation. Mathematically, for an expanded dataset $\mathcal{D}_{\mathrm{Opacity+}} := \mathcal{D}_{\mathrm{Opacity}} \cup \mathcal{D}_{\mathrm{Non-Opacity}}'$, where $\mathcal{D}_{\mathrm{Non-Opacity}}'$ is a random subset of $\mathcal{D}_{\mathrm{Non-Opacity}}$, each $X \in \mathcal{D}_{\mathrm{Opacity+}}$ has a $\mathrm{label} \in \{\text{Opacity}, \text{Non-Opacity}\}$, and the loss is defined as:
\begin{equation}
\resizebox{0.4\textwidth}{!}{$
\mathcal{L}_{bam} = \mathcal{A}\bigl(\mathcal{M}_{\mathrm{Non-Opacity}}\bigr)\,\mathds{1}\bigl(\mathrm{label} = \text{Non-Opacity}\bigr),
$}
\end{equation}
\noindent where $\mathcal{M}_{\mathrm{Non-Opacity}}$ is the activation mask for CXR $X$ in the non-opacity direction, and $\mathcal{A}(\cdot)$ returns the sum of the absolute values of elements in the input. $\mathds{1}(\cdot)$ represents the indicator function, which outputs $1$ if the condition is true and $0$ otherwise.
}
\subsection{Cross-Domain Alignment Module (CDAM)} After the model output passes through the Adaptive Activation Mask Selection Penalty Module, the generated CXRs for lung opacities can still be affected by inaccurate details and artifacts. {\color{black}While previous studies in image translation have predominantly relied on paired datasets, the challenge intensifies in unpaired settings where target domain guidance is less explicit. This complicates the learning of robust representations, potentially leading to generated images containing artifacts or hallucinated radiomic features that are absent in real chest X-rays (CXRs). To mitigate these issues and reduce the occurrence of Non-existent features in the model output, our approach incorporates an external prior, a pre-trained lung disease CXR classifier. This prior facilitates the alignment process by providing a structured guidance mechanism.} As depicted in the lower right section of \autoref{fig:Model_Overview} , a frozen pre-trained CXR disease multi-label classifier \cite{rajpurkar2017chexnet} is leveraged to complete this alignment protocol.

\paragraph{\textbf{Feature-Level Alignment (FA)}} We compute the $\ell_{2}$ distance between the latent visual feature maps extracted by the pretrained classifiers for the actual CXRs and the generated CXRs. Feature-level alignment provides guidance to promote the consistency of the invariant domain features. Given a disease classifier denoted as $ \mathcal{C} = \psi \circ \phi $, consisting of one feature extractor $\phi$ and one classifier head $\psi$, to obtain  $\text{Pred}$. Then, for the generated $\hat{X}_{\text {masked Non-Opacity}}$ and $X_{\text {Opacity}}$, the loss functions were computed as:\begin{equation}\mathcal{L}_{\text{structural}} = \left\| \text{IN}(\phi(\hat{X}_{\text{masked Non-Opacity}})) - \text{IN}(\phi(X_{\text{Opacity}})) \right\|^2,\end{equation}\begin{equation}\mathcal{L}_{\text{style}} = \left\| \text{GM}(\phi(\hat{X}_{\text{masked Non-Opacity}})) - \text{GM}(\phi(X_{\text{Opacity}})) \right\|^2,\end{equation}
where the total loss is then calculated as $\mathcal{L}_{feature} = \lambda_{style}\mathcal{L}_{\text{style}} + \mathcal{L}_{\text{structural}}$, we first perform Instance Norm (IN) \cite{isola2017image} and Gram Matrix (GM) to normalise the feature maps with the shape of $1024 \times 8 \times 8$ to extract the structural information and the style information, respectively. Then we set $\lambda_{style} = 0.5$ to control the weight of the $\mathcal{L}_{style}$ in the computation of $\mathcal{L}_{feature}$.

\paragraph{\textbf{Label Consistency Alignment (LCA)}}
{\color{black}
The classifier head delivers the prediction $\text{Pred} = \psi(\phi(x)) \in [0, 1]^{14}$ for 14 types of diseases depicted in \autoref{fig:Model_Overview}. Since only the opacity-related lung illnesses are of interest, only 6 of the 14 diseases and the positive label predictions of the CXRs from the source domain above the threshold  $p = 0.5$ are selected to compute the label alignment using a filter $\mathbf{q}$ consisting of 0 and 1. For a single $i$th CXR, this loss function is computed as:
\begin{equation}
\resizebox{0.4\textwidth}{!}{$
\mathcal{L}_{\mathrm{classifier}} = \ell_{\text{BCE}}\bigl(\hat{\text{Pred}}_{\text{inverted Non-Opacity}}, \text{Pred}_{\text{Opacity}}\bigr),
$}
\end{equation}
\noindent where $\ell_{\text{BCE}}$ represents the binary cross-entropy loss and  $\hat{\text{Pred}}_{\text{inverted Non-Opacity}} = f_{\text{inverted}}[\mathds{1}(\psi(\phi(\hat{X}_{\text{masked Non-Opacity}})) > p)] \cdot \mathbf{q}$,  $\text{Pred}_{\text{Opacity}} = \mathds{1}(\psi(\phi(X_{\text{Opacity}})) > p) \cdot \mathbf{q}$.
}
\subsection{Total Objective Function}
The total loss function is defined as:
\begin{align}
    \mathcal{L}_{total} &=  \lambda_{penalties}(\mathcal{P}_{upper} + \mathcal{P}_{repul} ) + \lambda_{bam} \mathcal{L}_{bam} \notag \\
    &\quad + \lambda_{feature}\mathcal{L}_{feature} + \lambda_{classifier}\mathcal{L}_{classifier} \notag \\
    &\quad + \lambda_{adv}\mathcal{L}_{adv} + \lambda_{rec}\mathcal{L}_{rec},
\end{align}
where $\mathcal{L}_{adv}$ is defined according to \cite{zhao2020unpaired} and $\mathcal{L}_{rec}$ represents the reconstruction loss using $\ell_{1}$, $\lambda_{penalties} = 0.01$, $\lambda_{bam} = 0.1$, $\lambda_{feature} = \lambda_{classifier} = 0.5$, $\lambda_{adv} = 2$, and $\lambda_{rec} = 1$ are set to train our translation model.

\section{Experimental Settings}
\subsection{Datasets}
\textbf{RSNA Opacity}: This dataset \cite{wang2017chestx} was created for the RSNA Kaggle Competition for pneumonia detection challenges \cite{shih2019augmenting}. It included three annotation CXR class labels: normal for 8851 images, lung opacity for 6021 pictures, and not normal for 11821 images. For our study, data were primarily used to train and evaluate the proposed method for the unpaired image translation model settings. The test set of the segmentation dataset of images is denoted as $\operatorname{RSNA}_{seg}$. 

\textbf{MIMIC-CXR-JPG \cite{johnson2019mimic}}: The MIMIC-CXR-JPG dataset was a collection of 377,110 CXR images. It comprised the CXRs, reports, and structured labels of 14 CXR observations extracted by CheXpert labeller \cite{irvin2019chexpert}. In our study, physicians manually selected ARDS patients for the evaluation of segmentation and classification performance. 1392 CXRs were used for ARDS classification, and 93 ARDS CXRs,$\operatorname{MIMIC}_{seg}$, were used for lung segmentation evaluation.

{\color{black}
\textbf{JSRT \cite{shiraishi2000development}}: JSRT is a widely used dataset in medical imaging research, consisting of 247 CXRs. Each image has a resolution of 2048 × 2048 pixels with 12-bit grayscale. The dataset includes gold-standard lung field masks. In our study, JSRT was primarily used to evaluate the quality of generated CXRs from the unpaired image translation model.
}

\subsection{Evaluation Metrics}
To evaluate the image quality of generated CXRs using our method, we employ Kernel Inception Distance\cite{binkowski2018demystifying} and Fréchet Inception Distance\cite{heusel2017gans} to qualify the visual similarity between the real CXRs and generated CXRs before and after the translation. Accuracy, precision, recall, and F1-scores  are used to qualify the performance of lesion detection in our binary classification problems. Segmentation performance is evaluated using mean Intersection over Union (mIoU) and sensitivity.

\subsection{Implementation Details and Baseline Methods} To train the proposed method in an adversarial framework, we optimised objectives for both the generators and discriminators, respectively, for $2.5 \times 10^{5}$ iterations with a batch size of 2. {\color{black} The input CXRs were first randomly cropped and then resized to 512 $\times$ 512 resolution.} We used the Adam optimizer \cite{kingma2014adam} with an initial learning rate of $1 \times 10 ^{-4}$ and $\beta$ values of 0.9 and 0.9999. A learning rate decay schedule was also applied every $1 \times 10^5$ iteration. The model was implemented in PyTorch Lightning using an RTX 3090 graphics card for both training and inference. To compare the unpaired image translation performance of our proposed method, we evaluated the translation results against other translation methods from \cite{torbunov2023uvcgan,lee2018diverse,zhu2017unpaired,huang2018multimodal,liu2017unsupervised}. To assess the segmentation quality and classification performance of the method's generated results, we used the fine-tuned CXR lung segmentation models \cite{zhao2017pyramid,karen2014very,gu2019net,ronneberger2015u}. The trained CXR disease detection model is based on YOLOv5 \cite{ultralytics2021yolov5} and other CXR classifiers with different backbones \cite{he2016deep, howard2017mobilenets, huang2017densely,tan2019efficientnet}.

\section{Discussion}

 \begin{table*}[!htb]
    \captionsetup{width=1\linewidth}
    \centering  
    \color{black}
    \caption{   \color{black}
        \textbf{Performance Evaluation of Unpaired Image Translation Models for CXR Lung Opacity Removal on the RSNA Dataset and Lung Segmentation Impact using $\operatorname{MIMIC}_{\text{seg}}$, $\operatorname{RSNA}_{\text{seg}}$, and $\operatorname{JSRT}_{\text{seg}}$ Datasets.}  The segmentation results are averaged across four segmentation methods (PSPNet, Unet, Cenet ,Vgg) on the same datasets. More details are provided in the \textit{Supplementary Material}. Metrics include FID, KID, mIoU $\Delta$, mIoU, and Sensitivity (in \%) on the RSNA, MIMIC-CXR ARDS, and JSRT test sets. Detailed results are provided in the supplementary material.
    }
    \resizebox{1\linewidth}{!}{ 
    \begin{tabular}{c|cc|ccc|ccc|ccc|}
    \toprule \toprule
    \multirow{2}{*}{\textbf{Model}} & 
    \multicolumn{2}{c|}{ \textbf{RSNA (Lung Opacity to Non-opacity)} } & 
    \multicolumn{3}{c}{$\operatorname{\textbf{MIMIC}}_{\text{seg}}$}  & 
    \multicolumn{3}{c}{$\operatorname{\textbf{RSNA}}_{\text{seg}}$} & 
    \multicolumn{3}{c}{$\operatorname{\textbf{JSRT}}_{\text{seg}}$} \\
    \cline{2-12} 
    & \textbf{Avg. FID $\downarrow$} & 
    \textbf{Avg. KID $\downarrow$} & 
    \textbf{mIoU $\Delta$} & 
    \textbf{mIoU $\uparrow$} &  
    \textbf{Sensitivity $\uparrow$} & 
    \textbf{mIoU $\Delta$} & 
    \textbf{mIoU $\uparrow$} &  
    \textbf{Sensitivity $\uparrow$} & 
    \textbf{mIoU $\Delta$} & 
    \textbf{mIoU $\uparrow$} &  
    \textbf{Sensitivity $\uparrow$} \\
    \toprule
    
    Original CXRs & 81.8 & 0.043 (0.0104) &  & \textbf{86.66}  & 91.16 &  & 73.19 & 79.56 &  & \textbf{ 91.15} & 97.55  \\
    \midrule
    Munit \cite{huang2018multimodal} & 109.4 & 0.073 (0.0038) & -14.59  & 72.07  & 86.85  & -10.61 & 62.58 & 77.03  & -5.55  &  85.60 & 95.04  \\
    Unit \cite{liu2017unsupervised} & 103.2 & 0.061 (0.0016) & -11.94 &  74.72 
    &  83.88  & -5.73  & 67.46  & 75.92  & -4.55  &  86.60 & 95.15  \\
    CycleGAN \cite{zhu2017unpaired} & 208.3 & 0.216 (0.0068) & -26.93  & 59.73  & 66.72   & -14.35  & 58.84   & 65.39  & -10.29  &  80.86 & 88.42   \\
    Uvcgan \cite{torbunov2023uvcgan} & 210.4 & 0.225 (0.0065) &  -5.14  & 81.52 & 86.93  & -6.53  &  66.66 & 72.97  & -2.11 &  89.04 & 95.76  \\
    Drit \cite{lee2018diverse}& 117.6 & 0.087 (0.0041) & -10.68  & 75.98  & 83.52  & -6.42  & 66.77  &  75.88  &  -2.5 &  88.65 & 96.18  \\
    \midrule
    \rowcolor[gray]{0.5} \textbf{Ours} & \textbf{67.18} & \textbf{0.01604 (0.0034)} & \textbf{-0.46 } &  86.20  & \textbf{92.68 }  & \textbf{+3.39} & \textbf{76.58} & \textbf{85.58} & \textbf{-0.07 } & 91.08  &  \textbf{97.62}   \\
    \bottomrule
    \bottomrule
    \end{tabular}  
    }
    \label{tab:baseline_method_comparsion}
    \end{table*}

\begin{figure}[!htb]
    \centering
    \captionsetup{width=1\linewidth}
    \includegraphics[width=1\linewidth]{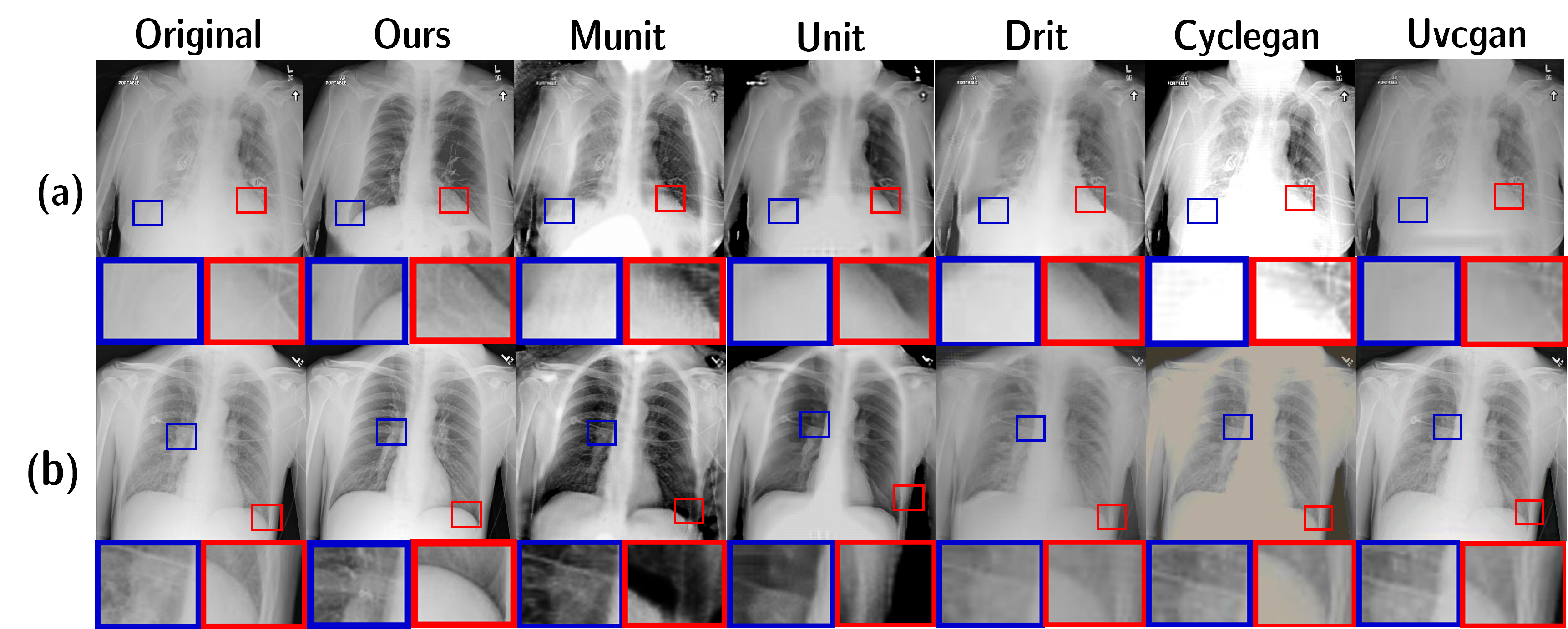}
    \caption{ \textbf{Qualitative Comparison of Translated CXRs Generated by Various Methods.} Each row (a) through (c) represents two examples of CXR. Columns from left to right show the original CXR, our method, Munit, Unit, Drit, CycleGAN, and Uvcgan. CXRs without lung opacities generated by our method show better-translated image quality in terms of appropriate pixel intensity, clear lung borders, and the preservation of necessary details compared to the other methods.}
    \label{fig:baseline_methods_comparsion}
\end{figure}

\paragraph{\textbf{Overall Performance and Ablation Studies}}
With the proposed modules, the results shown in the first two columns of  \autoref{tab:baseline_method_comparsion} indicate superior performance in terms of FID and KID metrics compared to current methods for unpaired image translation.  As demonstrated, the proposed method achieves the lowest average FID and KID compared to other unpaired image translation methods for unpaired Non-opacity CXR translation. The qualitative comparison across the methods is displayed in \autoref{fig:baseline_methods_comparsion}. For the ablation studies,    the contribution of each component to the performance metrics is displayed in \autoref{table: ablation_studies}.

\begin{figure}[!htb]
    \centering
    \captionsetup{width=1\linewidth}
    \includegraphics[width=1\linewidth]{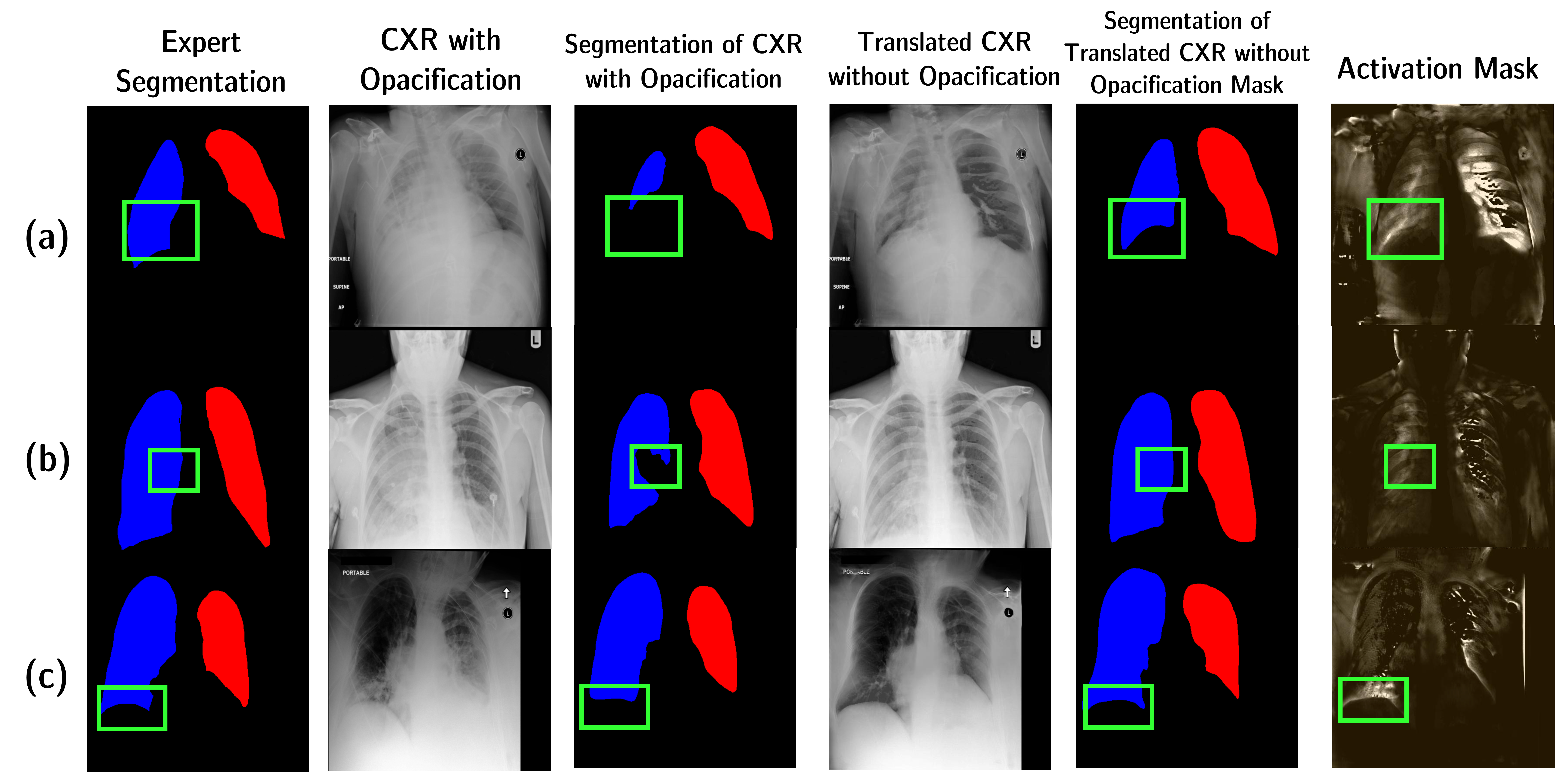}
    \caption{\textbf{Qualitative Comparison of Segmentation Results on Lung CXRs with Opacities and Translated Lung CXRs without Opacities.} Each row (a) through (c) represents three different CXR images. Columns from left to right show: Expert Segmentation Mask, CXR with Opacification, Segmentation of CXR with Opacification, Translated CXR without Opacification, and Segmentation of Translated CXR without Opacification. After the translation, the regions of lung opacities are removed, enabling more accurate lung segmentations given the same lung segmentation model.}
    \label{fig:segmentation_results}
\end{figure}

\paragraph{\textbf{Semantic Consistency in Transformed Results}} 
In \autoref{tab:baseline_method_comparsion}, we compared the lung segmentation results on subsets of the MIMIC and RSNA datasets. On the test set of the RSNA dataset, we focused on examining the performance in lung segmentation of regions estimated to have no lung opacity. The segmentation results on $\operatorname{RSNA}_{seg}$ indicated a 3.39\% gain in mIoU and an 6\% improvement in segmentation sensitivity compared to other methods which struggled to preserve the semantic information. Other image translation methods do not consider the original features when using the activation mechanism for blending, which makes these methods vulnerable to losing anatomical details after translation. The activation masks produced by the generator reflect this insight, corresponding to the regions revealed in the activation mask after translation. As shown in $\operatorname{MIMIC}_{seg}$ of \autoref{tab:baseline_method_comparsion}, for the more challenging CXR translation in ICU ARDS patients whose symptoms often involve extensive lung opacity and where the ICU devices obscure the lung regions, our proposed method can preserve the semantic disease pattern in the translation compared to those of other methods with only less than 1\% segmentation degradation in mIoU.

\begin{figure}[!htb]
\centering

\includegraphics[width=0.90\linewidth]{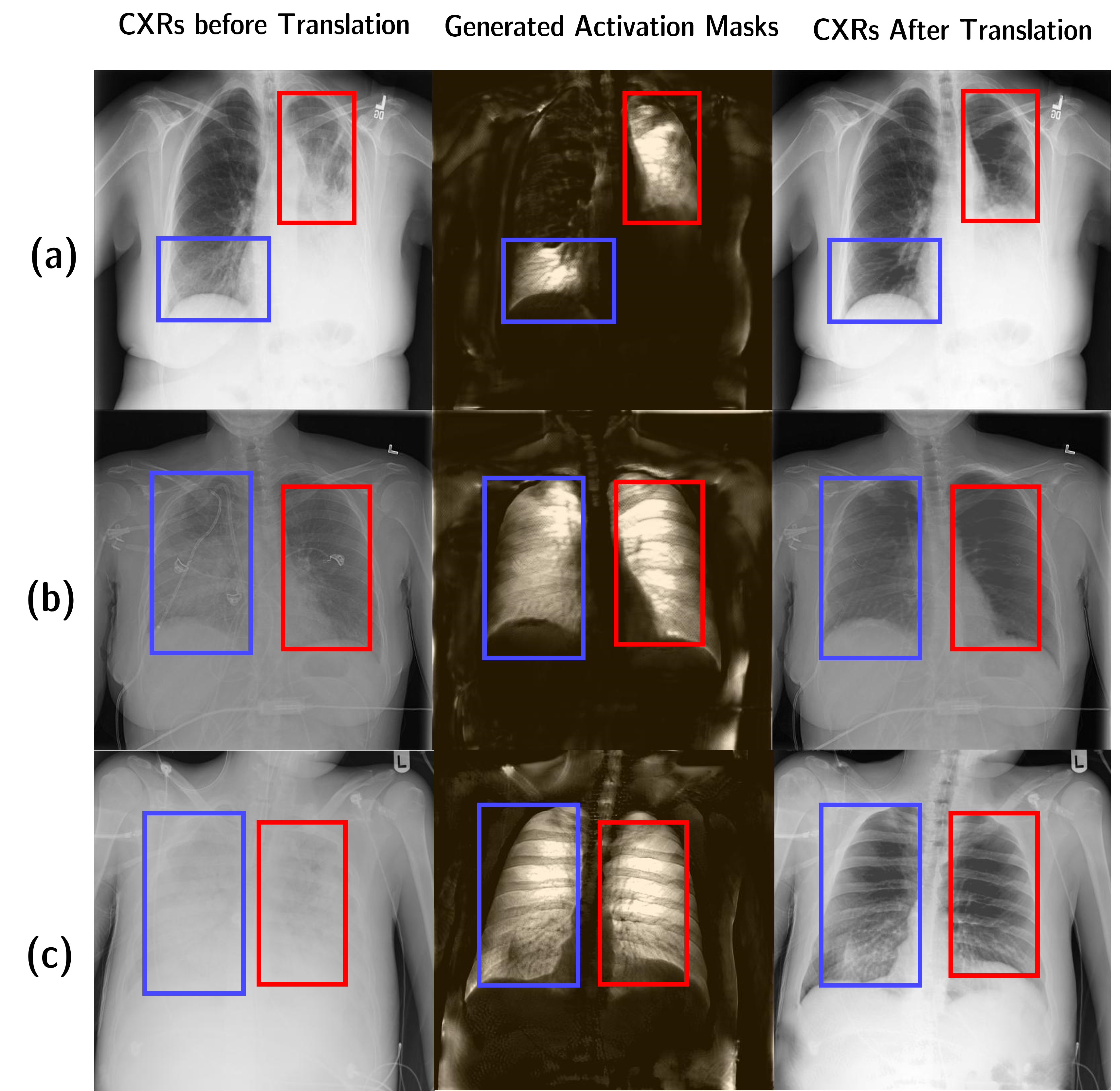}
\caption{ \color{black}\textbf{Comparative Analysis of Translation Results and Activation Masks.} Rows (a) through (c) present distinct chest X-ray (CXR) examples. Activation masks generated by the model correctly delineate regions of transformation during translation, providing a mechanism to interpret the process of translation during the removal of lung opacities.}
\label{fig:activation_mask_interpretation}
\end{figure} 

\paragraph{\textbf{Classification of Lung Opacities with Translated CXR Images}} To examine the potential of the model-generated CXRs and activation masks for training lung opacity classifiers, we designed an experiment to create a synthetic dataset from both domains to train a binary CXR classifier for lung opacity detection. Further details are in \textit{Supplementary Material}. We compared the classification performance of the classifiers to those purely trained on real CXRs from both domains. {\color{black} The purpose of this was to demonstrate that the translated CXRs without lung opacities are visually accurate and realistic while preserving relevant anatomical features
for lesion classification. } In \autoref{tab:classification_results_comparison}, we present the CXR classification performance for both lung opacities and ARDS, trained on synthesized CXR data, in terms of both F1 score and accuracy.
{\color{black}The results demonstrate that the proposed method can effectively retain the valuable anatomical features present in CXRs, even after translation to the target domain.}

\begin{table}[!htb]
\centering
\color{black}
\caption{\color{black}\textbf{Classification Performance Comparison on RSNA and MIMIC-ARDS Datasets with Classifiers Trained on Translated CXRs and Activation Masks Produced by Our Method.} Accuracy (Acc.), F1-score, Recall, and Precision are evaluated. ``Original" indicates real CXRs, ``Generated" refers to our method's translated CXRs, and ``Mask" represents the generated activation masks. Two tasks are compared: 1) Real CXRs vs. translated CXRs for RSNA lung opacity classification, and 2) Real CXRs vs. activation masks for both RSNA lung opacity and MIMIC-ARDS patient CXR classification. \textbf{Bold} text indicates the highest scores in each metric.}
\resizebox{\linewidth}{!}{
\begin{tabular}{ccccccccccc}
\toprule \toprule
\multirow{3}{*}{\textbf{Model}} & \multirow{3}{*}{\textbf{Data Type}} & \multicolumn{4}{c|}{\textbf{RSNA}} & \multicolumn{4}{c}{\textbf{MIMIC-ARDS}} \\
\cline{3-10}
& & \textbf{Acc.} & \textbf{F1} & \textbf{Recall} & \textbf{Precision} & \textbf{Acc.} & \textbf{F1} & \textbf{Recall} & \textbf{Precision} \\
\midrule
\multirow{3}{*}{Resnet50}
& Original & 0.8483 & 0.7367 & 0.7799 & 0.6982 & 0.8467 & 0.7652 & 0.8742 & 0.6803 \\
& Generated & 0.7978 & 0.8755 & 0.8504 & 0.9020 & - & - & - & - \\
& Mask & \textbf{0.8803} & \textbf{0.9580} & \textbf{0.9765} & \textbf{0.9402} & \textbf{0.8489} & \textbf{0.8452} & \textbf{0.9154} & \textbf{0.7852} \\
\midrule
\multirow{3}{*}{Mobilenet}
& Original & \textbf{0.878} & 0.8470 & 0.9164 & 0.7874 & 0.8133 & 0.6573 & 0.644 & 0.6712 \\
& Generated & 0.7961 & 0.8851 & 0.7953 & \textbf{0.9973} & - & - & - & - \\
& Mask & 0.8535 & \textbf{0.9436} & \textbf{0.9599} & 0.9278  & \textbf{0.8267} & \textbf{0.7283} & \textbf{0.7429}  &  \textbf{0.7142} \\
\midrule
\multirow{3}{*}{Efficientnet}
& Original & \textbf{0.9093} & 0.8864 & 0.9260 & 0.8501 & 0.7733 & 0.4684 & 0.4024 & 0.5603 \\
& Generated & 0.8174 & 0.8921 & 0.8320 & 0.9616 & - & - & - & - \\
& Mask & 0.8635 & \textbf{0.9519} & \textbf{0.9592} & \textbf{0.9447}   & \textbf{0.7933} & \textbf{0.7497} & \textbf{0.8002} & \textbf{0.7052} \\
\midrule
\multirow{3}{*}{Densenet}
& Original & 0.8763 & 0.8642 & 0.8220 & 0.9110 & \textbf{0.8867} & 0.8071 & 0.8445 & 0.7728 \\
& Generated & 0.7837 & 0.8789 & 0.7852 & \textbf{0.9979} & - & - & - & - \\
& Mask & \textbf{0.8866} & \textbf{0.9671} & \textbf{0.9614} & 0.9729   & 0.8267 & \textbf{0.8396} & \textbf{0.9021} & \textbf{0.7852} \\
\bottomrule
\bottomrule 
\end{tabular}
}
\label{tab:classification_results_comparison}
\end{table}

% \begin{figure}[!htb]
%     \centering
%     \includegraphics[width=1\linewidth]{ablation_studiesv2.png}
%     \caption{\textbf{Qualitative comparison of model components in ablation studies.} Each row represents a different CXR sample, while columns show the effects of incrementally adding model components from left to right: AASPM (Adaptive Activation Selection Penalty Module), FA (Feature Alignment), LCA (Label Consistency Alignment), and BAML (Bidirectional Feature Minimisation Loss). FA + LCA produce the overall visual appearance guidance and high-level alignment of the disease types, while the AASPM is better handling the lung borders and decision of determining features carried from the initial CXRs in final output to preserve the necessary anatomical details.}
%     \label{fig:ablation_studies}
% \end{figure}

\begin{table}[!htb]
\centering
\caption{\textbf{Ablation Studies of Model Components Evaluated on the RSNA Test Set.} Components include: AASPM (Adaptive Activation Mask Selection Penalty Module), FA (Feature Alignment), LCA (Label Consistency Alignment), and BAML (Bidirectional Activation Minimisation Loss). Avg. FID and Avg. KID represent average Fréchet Inception Distance and Kernel Inception Distance respectively, with lower values ($\downarrow$) indicating better performance.}
\resizebox{0.9\linewidth}{!}{
\begin{tabular}{cccc|cc}
\toprule \toprule

 \textbf{AASPM} & \textbf{FA} & \textbf{LCA} & \textbf{BAML} & \textbf{Avg. FID $\downarrow$}  & \textbf{Avg. KID $\downarrow$} \\
\midrule
  \xmark & \xmark & \xmark & \xmark & 77.05845  & 0.03214 \\
  \cmark & \xmark & \xmark & \xmark & 74.19213 &  0.02471 \\
\xmark & \cmark & \xmark & \xmark & 73.94616 & 0.02427 \\
\cmark & \cmark & \xmark & \xmark & 70.05348 & 0.01891 \\
\Xhline{1\arrayrulewidth}
 \cmark & \cmark & \cmark & \xmark & 68.73514& 0.01612 \\
\cmark & \cmark & \cmark & \cmark & 67.17715 & 0.01604 
  \\
\bottomrule \bottomrule
\end{tabular}
}

\label{table: ablation_studies}
\end{table}

\begin{figure}[!htb]
    \centering
    \includegraphics[width=1\linewidth]{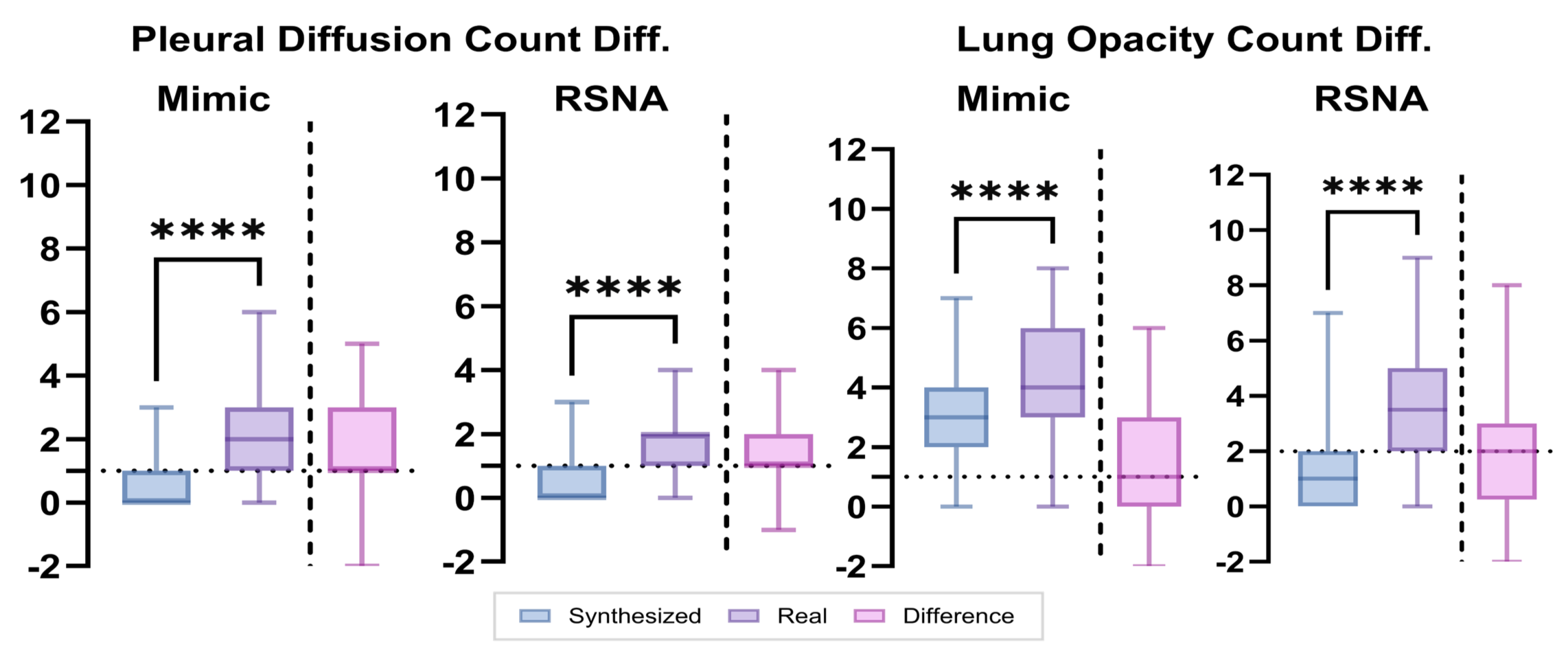}
    \caption{
        \textbf{Box plots of the count of pathological regions in real and translated chest X-rays (CXRs) from the RSNA and MIMIC-CXR test sets.}  
    }
    \label{fig:object_detection_plot_v1}
\end{figure}

\paragraph{\textbf{Quantitative Assessment of Lung Opacity Detection Using CXR Lesion Detection Model}} To determine whether the number of the lung opacity regions in the original CXR is reduced after the translation, we utilised an object detector based on YOLOv5 \cite{ultralytics2021yolov5} to perform lesion detection on lung opacity in both the original and generated CXRs. In \autoref{fig:object_detection_plot_v1}, our evaluation of lung opacity detection tasks on the generated CXRs demonstrated that the proposed method effectively reduced the number of lung opacities and pleural effusions observed in CXRs after translation, with a $p$-value $\leq 0.05$ for both the RSNA and MIMIC-CXR datasets.

{\color{black}\paragraph{\textbf{Discriminative Power and Interpretability of Activation Masks}} We validated the capability of the activation masks to interpret disease patterns in lung opacity classification. The activation masks were designed to assist the translation in generating the final output of CXRs. We hypothesised that the activation masks can explain lung opacity conditions while completing this role. To this end,  we followed similar experimental settings for the classification of lung opacities. Instead of using generated CXR images to train CXR disease classifiers, we used the activation masks as the training data source. As shown in the ``mask" rows of \autoref{tab:classification_results_comparison}, the results suggest that the activation masks produced by our method provide discriminative features to differentiate outputs in lung opacity detection. In the more challenging scenario of ARDS CXR classification, a similar trajectory of improvement was observed. Qualitatively, examples in \autoref{fig:activation_mask_interpretation} corroborate these findings, illustrating that the activation masks offer a way to visualise the relevant areas and help explain the model behaviour  during the translation. }

\section{Conclusion}
In this study, we proposed an unpaired image translation model with adaptive activation mask selection and a cross-domain alignment module to remove lung opacity in CXRs. Through the analysis of activation masks, we enhanced mask learning by incorporating modified penalties. The proposed approach accurately identifies lung opacity regions in CXRs that require attention during translation and provides a means to assess the interpretability of the translation process. Furthermore, the cross-alignment module helped acquire visual cues and align anatomical shapes of the lungs in real CXRs under unpaired settings. Our research highlights that in unpaired image translation of CXR lung opacities, both the translation and segmentation results demonstrate superior performance compared to other approaches, particularly on the RSNA and MIMIC-CXR ARDS datasets. Empirical validation of classification performance shows that classifiers trained on our model-generated activation masks outperform those trained on original CXR images in identifying lung opacity.  {\color{black} In the future, we aim to minimize visual distortions and enhance the control of activation mask learning, especially in challenging cases with weak visual cues. This improvement is expected to broaden the applicability of our CXR translation method to previously unseen settings. After further optimization, we will conduct additional testing on out-of-distribution data through prospective evaluations and expert reviews to verify its reliability and potential for medical applications. }

\section*{Declaration of Competing Interest}
The authors declare that they have no known competing financial interests or personal relationships that could have potentially appeared to influence the work reported in this paper.

\section*{Acknowledgments} 
\noindent This study was supported in part by the ERC IMI (101005122), the H2020 (952172), the MRC (MC/PC/21013), the Royal Society (IEC\textbackslash NSFC\textbackslash 211235), the NVIDIA Academic Hardware Grant Program, the SABER project supported by Boehringer Ingelheim Ltd, NIHR Imperial Biomedical Research Centre (RDA01), The Wellcome Leap Dynamic resilience program (co-funded by Temasek Trust)., UKRI guarantee funding for Horizon Europe MSCA Postdoctoral Fellowships (EP/Z002206/1), UKRI MRC Research Grant, TFS Research Grants (MR/U506710/1), and the UKRI Future Leaders Fellowship (MR/V023799/1).
Dominic Marshall is supported by an MRC clinical research training fellowship (award MR/Y000404/1) and the Mittal Fund at Cleveland Clinic Philanthropy .

%% \section{}
%% \label{}

%% For citations use: 
%%       \citet{<label>} ==> Jones et al. [21]
%%       \citep{<label>} ==> [21]
%%

%% If you have bibdatabase file and want bibtex to generate the
%% bibitems, please use
%%
%%  \bibliographystyle{elsarticle-num-names} 
%%  \bibliography{<your bibdatabase>}

%% else use the following coding to input the bibitems directly in the
%% TeX file.

%% The Appendices part is started with the command \appendix;
%% appendix sections are then done as normal sections

% \newpage
\bibliographystyle{elsarticle-num-names} 
\bibliography{references.bib}

% \begin{thebibliography}{00}

% %% \bibitem[Author(year)]{label}
% %% Text of bibliographic item

% \bibitem[ ()]{}

% \end{thebibliography}

\end{document}